# Optimizing Decoy State Enabled Quantum Key Distribution Systems to Maximize Quantum Throughput and Detect Photon Number Splitting Attacks with High Confidence

Logan O. Mailloux, *Member, IEEE*, Michael R. Grimaila, *Senior Member, IEEE*,
Douglas D. Hodson, Ryan Engle, *Member, IEEE*, Colin McLaughlin, and Gerald Baumgartner

*Abstract*—Quantum Key Distribution (QKD) is an innovative quantum communications protocol which exploits the laws of quantum mechanics to generate unconditionally secure cryptographic keying material between two geographically separated parties. The unique nature of QKD shows promise for high-security applications such as those found in banking, government, and military environments. However, QKD systems contain implementation non-idealities which can negatively impact their performance and security. In particular, QKD systems often employ the decoy state protocol to improve system throughput and mitigate the threat of Photon Number Splitting (PNS) attacks. In this work, a detailed analysis of the decoy state protocol is conducted which optimizes both performance in terms of quantum throughput and security with respect to detecting PNS attacks. The results of this study uniquely demonstrate that the decoy state protocol can ensure PNS attacks are detected with high confidence, while maximizing the system's secure key generation rate at no additional cost. Additionally, implementation security guidance is provided for QKD system developers and users.

*Index Terms*—Quantum Key Distribution, Decoy State Protocol, Photon Number Splitting Attack, Implementation Security

## I. Introduction

Quantum Key Distribution (QKD) is a revolutionary quantum communications protocol which provides the means for two geographically separated parties to generate unlimited amounts of unconditionally secure symmetric keying material. Unlike conventional key distribution techniques, the security of QKD systems rests on the laws of quantum mechanics and not on computational complexity [1]. In theory, these attributes make QKD well suited for high-security applications such as banking, government, and military environments. However, QKD is a nascent technology with implementation non-idealities and practical engineering limitations which can negatively impact the system's performance and security [2]. For example, in order to mitigate implementation vulnerabilities commercially viable QKD systems (i.e., those which balance cost, performance, and security towards affordability [3]) often employ the decoy state protocol to detect Photon Number Splitting (PNS) attacks and improve system performance [4].

While the decoy state protocol is well studied with respect to unconditionally secure key generation (see background for details), this work uniquely examines the protocol's ability to both detect PNS attacks and maximize quantum throughput. The results of this study: (i) provide an optimization of the decoy state protocol which maximizes secret key generation rates and the protocol's ability to detect PNS attacks with high confidence; (ii) demonstrate the effectiveness of the optimization to detect PNS attacks; (iii) offer implementation security engineering considerations for QKD system designers and users; and (iv) present a repeatable methodology for studying quantum communication issues related to non-ideal protocol implementations.

This article is organized as follows: first an introduction to QKD is provided with an emphasis on security vulnerabilities, the PNS attack, and the decoy state protocol. In Section III, the research method is explained, including a comprehensive listing of decoy state enabled QKD systems. Section IV details the decoy state protocol's ability to detect PNS attacks across 40 operationally relevant decoy state protocol configurations. Based on these results, an optimization of the protocol is presented and demonstrated along with implementation security recommendations. Lastly, conclusions and future work are discussed in Section V. For security specialists desiring to further understand QKD, please see [5], [6], [7]. For comprehensive physics-based reviews, please see [1], [4].

Manuscript received June 15, 2016. This work was supported by the Laboratory for Telecommunication Sciences [grant number 5743400-304-6448]. This work was supported in part by a grant of computer time from the DoD High Performance Computing Modernization Program at the Air Force Research Laboratory, Wright-Patterson AFB, OH.

L. O. Mailloux; M. R. Grimaila; D. D. Hodson; and R. D. Engle are with the Air Force Institute of Technology, Wright-Patterson AFB, OH 45433-7765 (email: {logan.mailloux; michael.grimaila; douglas.hodson; ryan.engle}@afit.edu).

C. McLaughlin is a Research Physicist at the Naval Research Laboratory, Washington, D.C. 20375 (email: colin.mclaughlin@nrl.gov).

G. Baumgartner is a Research Physicist at the Laboratory for Telecommunication Sciences, College Park, MD 20740 (email: gbaumgartner@ltsnet.net).

## II. Quantum Key Distribution (QKD)

The genesis of QKD traces back to the late 1960s, when Wiesner first proposed the idea of encoding information on polarized photons using two conjugate bases [8]. In 1984, Bennett and Brassard extended this idea by introducing the first QKD protocol, known as "BB84," to generate shared secret keying material between two parties [9]. Today, QKD is gaining attention as an important development in the cybersecurity solution space because of its ability to generate unlimited amounts of symmetric keying material for use with the One-Time-Pad (OTP) – the only known encryption algorithm to achieve perfect secrecy [10], [11]. In this way, QKD enables unbreakable communications and has inspired research efforts across Asia, Europe, and North America [12].

### A. The BB84 QKD Protocol

While there are many competing QKD protocols, BB84 is primarily considered in this work because it remains a popular implementation choice and is relatively easy to understand [1].

Fig. 1 illustrates a notional QKD system configured to securely generate the secure shared key $K$, which is used to encrypt/decrypt sensitive data, voice, or video communications. The QKD system consists of a sender "Alice," a receiver "Bob," a quantum channel (i.e., an optical fiber or direct line of sight free space path), and a classical channel (i.e., a conventional networked connection). Alice is shown with a laser source configured to generate and prepare single photons, known as quantum bits or "qubits." The encoded photons are then transmitted over the quantum channel to Bob, whom measures them using specialized single photon detectors. This exchange of encoded single photons is described by the BB84 protocol.

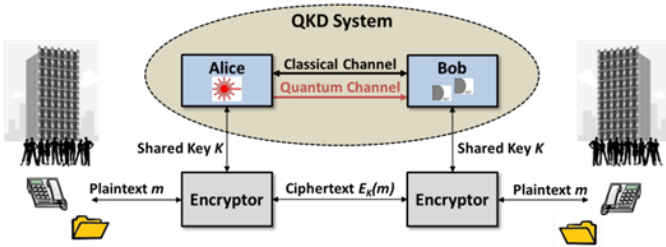

Fig. 1. A Quantum Key Distribution (QKD) system context diagram. The sender "Alice" and receiver "Bob" generate shared secret key $K$ for use in data encryption/decryption.

Table I describes the BB84 protocol as a prepare and measure protocol where Alice encodes photons in one of four polarization states (e.g., ↔, ↕, ↗, or ↘) according to a randomly selected bit value (0 or 1) and basis (⊕ for the pair ↔,↕ or ⊗ for the pair ↗,↘). Once Alice randomly prepares the photons, they are sent to Bob where he measures each photon using a randomly selected basis (⊕ or ⊗). If Alice's encoding and Bob's decoding bases match, the photon's bit value is read correctly with a high probability. Otherwise a random result occurs (i.e., equal likelihood of a 0 or 1). This is due to the inherent uncertainty in the measurement of an unknown (i.e., a randomly encoded) single photon.

More specifically, the security of the BB84 QKD quantum communications protocol is based on uncertainty of the measurement result when using two conjugate bases (i.e., ⊕ or ⊗) to randomly prepare and measure photons [13]. For example, during an intercept-resend attack anyone attempting to listen on the quantum channel must randomly select a measurement basis and will necessarily introduce detectable errors. This will increase the protocol's measured Quantum Bit Error Rate (QBER) and if the QBER ever exceeds the protocol's security threshold (e.g., QBER > 11% [4]), the secret key distribution process is aborted (or restarted) as it is assumed an eavesdropper is active on the quantum channel.

TABLE I
EXAMPLE BB84 PROTOCOL

| Alice Prepares | | | Bob Measures | |
|---|---|---|---|---|
| Bit | Basis | Prepared State | Basis | Result |
| 0 | ⊕ | \|↔⟩ | ⊕ | 0 |
| 1 | ⊕ | \|↕⟩ | ⊕ | 1 |
| 0 | ⊕ | \|↔⟩ | ⊗ | random |
| 1 | ⊕ | \|↕⟩ | ⊗ | random |
| 0 | ⊗ | \|↗⟩ | ⊕ | random |
| 1 | ⊗ | \|↘⟩ | ⊕ | random |
| 0 | ⊗ | \|↗⟩ | ⊗ | 0 |
| 1 | ⊗ | \|↘⟩ | ⊗ | 1 |

### B. Vulnerabilities in Protocol Implementation

BB84 security proofs assume several idealities, including perfect on-demand single photon sources, lossless quantum transmission, perfect transmitter-receiver basis alignment, and perfect single photon detection [14]. However, these security assumptions are not valid when building real-world systems which deviate from theoretical protocols [2]. For example, reliable on-demand single photon sources are not currently available nor are they expected in the near term [1]. Therefore, most QKD systems attenuate classical laser pulses down from millions of photons to weak coherent pulses with an average photon number less than one. More specifically, the number of photons contained in the pulse is represented using a Poisson distribution with a low (i.e., <1) Mean Photon Number (MPN)

$$P(n|\mu) = \frac{\mu^n e^{-\mu}}{n!} \quad (1)$$

where $\mu$ is the average number of photons in a pulse (i.e., $\mu$ is the MPN) and $n$ represents the number of photons in the pulse (i.e., $n = 0, 1, 2, 3, ..., N$). For example, with a typical MPN, $\mu = 0.5$, nearly 60% of the pulses have zero photons, 30% of the pulses have one photon, and 9% of the pulses have two or more photons. This means nearly 23% of the non-empty pulses emitted by Alice are non-ideal multiphoton pulses which leak information about the "unconditionally secure" QKD-generated secret key to eavesdroppers. This introduces a significant security vulnerability into the QKD protocol.

### C. Photon Number Splitting (PNS) Attacks

The PNS attack is a powerful attack designed to take advantage of the multiphoton vulnerability in order to obtain a full copy of Alice and Bob's shared secret key bits without introducing errors and thus increasing the QBER [15], [16]. A brief introduction to the PNS attack is given here, with a detailed, yet easily understandable engineering-oriented explanation available in [17].

Fig. 2 provides a simplified depiction of the eavesdropper "Eve" conducting a PNS attack against the QKD system (i.e., Alice and Bob). In accordance with QKD security proofs, Eve is an all-powerful adversary limited only by the laws of quantum mechanics [4]. She is allowed full control of the quantum channel to introduce losses or errors and may eavesdrop on, but not fabricate, messages exchanged on the classical channel. In order to conduct the PNS attack, Eve replaces the quantum channel with a quantum teleportation channel which enables the lossless transmission of photons from Alice to Bob using the properties of entangled quantum systems [18]. In order to avoid obvious disclosure, a geographically separated Eve′ entity is also required to regulate the lossless transmission of photons as to not exceed Bob's expected detection rate.

For each pulse Alice generates, Eve performs a specialized Quantum Non-Demolition (QND) measurement to determine the number of photons in each pulse $n = 0, 1, 2, 3, ..., N$ [19]. If $n \leq 1$, Eve blocks the pulse and sends nothing to Bob. If $n \geq 2$, Eve splits one photon from the pulse and stores it in her quantum memory. She then quantum teleports the remaining $n - 1$ photons to Bob. This attack scheme allows Eve to store an identical encoded copy of each photon sent to Bob without introducing additional errors (which are typically used for detecting eavesdroppers). Once Alice and Bob complete their quantum exchange, they must announce measurement basis information over the classical channel where Eve is able to listen. Eve can then correctly measure each stored photon, and thus, obtain a complete copy of the QKD-generated "secure" key bits.

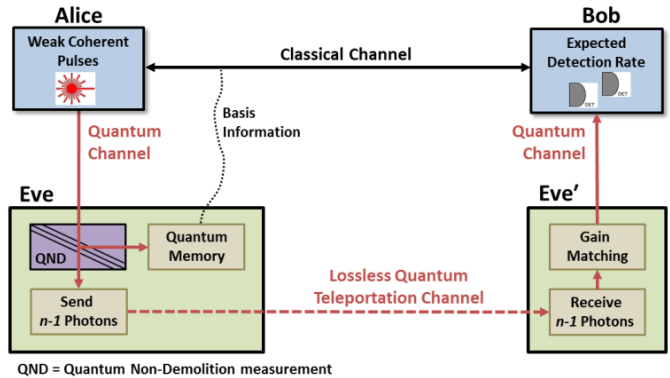

Fig. 2. The eavesdropper (Eve and Eve′) is shown conducting a Photon Number Splitting (PNS) attack against the QKD system (Alice and Bob).

### D. The Decoy State Protocol

In 2003, the decoy state protocol was introduced to detect PNS attacks [20]. It was quickly improved upon in a series of works [21], [22], [23], [24], [25]; and is now widely employed in commercially viable QKD systems such as Toshiba's record holding system [26] and the world's largest QKD network [27]. In particular, the decoy state protocol is advantageous as it is relatively easy to implement (low cost), increases the system's distributed secret key rate (high performance), and mitigates the PNS attack (high security).

As described by Ma et al., the decoy state protocol extends the BB84 protocol by configuring Alice to randomly transmit three types of pulses: 1. Signal; 2. Decoy; and 3. Vacuum as described in Table II [22]. Thus, Alice randomly generates signal, decoy, and vacuum pulses according to their prescribed occurrence percentages and respective MPNs where the state of each pulse must be indistinguishable to Eve (i.e., identical pulse shape, wavelength, duration, etc.) in order to maintain integrity of the security protocol. Eve cannot know *a priori* the type of pulse received during quantum exchange, the only information available to her is each pulse's specific number of photons $n = 0, 1, 2, 3, ..., N$ which she determines using her QND measurement.

TABLE II
EXAMPLE DECOY STATE PROTOCOL CONFIGURATION

| State | Purpose | MPN | Occurrence Percentage |
|---|---|---|---|
| Signal $\mu$ | The signal state is used to generate secret key and facilitates improved performance by using a higher MPN (i.e., 0.5 is greater than the value 0.1 typically employed in non-decoy state protocol QKD systems). | 0.5 | 70% |
| Decoy $\nu$ | The decoy state is used to increase the likelihood of detecting unauthorized eavesdropping on the quantum channel through statistical differential analysis with the signal state. | 0.1 | 20% |
| Vacuum $Y_0$ | The vacuum state is used to determine the noise on the quantum channel known as the "dark count" (i.e., detections when no photons are sent). | 0.0 | 10% |

### E. Unconditionally Secure Key Generation

While the decoy state protocol was introduced to detect PNS attacks, to date it has been primarily used to increase unconditionally secure key rates. More specifically, decoy state research has focused on understanding and bounding Alice's single photon generation, $Q_1$, as it pertains to secret key generation $R$ [22]

$$R \geq q\{Q_1[1 - H_2(e_1)] - Q_\mu f(E_\mu) H_2(E_\mu)\} \quad (2)$$

where $q$ is the protocol efficiency (e.g., <1), $Q_1$ is the estimated single photon contribution, $e_1$ is the estimated error rate of single photon detections, $Q_\mu$ is the signal state gain, $E_\mu$ is the signal state QBER, $f(E_\mu)$ is the error reconciliation efficiency, and $H(\{E_\mu, e_1\})$ is Shannon's binary information function [11]. This additional complexity is necessary because only pulses emitted by Alice containing a single photon (i.e., $Q_1$), those known as "tagged bits," can contribute to the QKD-generated unconditionally secure key $R$ [22].

In their 2005 work, Ma et al. optimized the signal state MPN ($\mu = 0.5$) to maximize the number of single photon pulses generated by Alice (i.e., $Q_1$) [22]. Following this seminal work, many others have studied the single photon bound $Q_1$ to account for fluctuations in laser sources [28], [29], [30], [31], [32], [33], [34], [35], [36], [37], [38], [39] and finite key size statistics [25], [40], [41], [42]. In addition to these theoretically focused works, several practically-oriented experimental demonstrations have been accomplished as detailed in Table III (discussed in Section III).

*F. Detecting PNS Attacks*

The decoy state protocol is designed to detect PNS attacks by comparing the signal and decoy states during quantum exchange, and specifically, the photon number dependent yields of the signal state $Y_n^{signal}$ and the decoy state $Y_n^{decoy}$ are examined with the security condition [21]

$$Y_n^{signal} = Y_n^{decoy} \qquad (3)$$

where $Y_n$ (signal or decoy) represents the conditional probability that Bob detects a pulse given Alice sent an $n$-photon pulse. Formally, $Y_n = Y_0 + \eta_n - Y_0\eta_n \cong Y_0 + \eta_n$ where $Y_0$ is the measured dark count rate and $\eta_n = 1 - (1-\eta)^n$ is the photon number specific efficiency based on the number of photons, $n$, in each pulse and the measured quantum efficiency $\eta$ when treating each photon independently. Lastly, the joint probability $Y_0\eta_n$ is disregarded because it is insignificant compared to $Y_0$ and $\eta_n$.

Under normal operational conditions (i.e., when no PNS attacks are occurring), the signal and decoy state yields should be the same for each $n$-photon yield regardless of its state. For example, $Y_1^{signal} = Y_1^{decoy}$ should always be true for a given QKD architecture because the signal and decoy state yields (i.e., $Y_n = Y_0 + \eta_n$) are primarily based on fixed quantum efficiencies and not the state type. If ever $Y_1^{signal} \neq Y_1^{decoy}$, an eavesdropper is assumed to be actively listening on the key distribution channel and the secret key is assumed compromised. While [21] also proposes an error-based condition $e_n^{signal} = e_n^{decoy}$, it is not considered in this work since the PNS attack does not introduce additional errors [15].

### III. Research Methodology

Table III provides a chronological listing of practically-oriented decoy state protocol experiments. From this comprehensive survey, there is relatively little consistency amongst protocol configurations as signal state MPNs range from 0.27 to 0.80 and decoy state MPNs range from 0.08 to 0.20. Additionally, it is worthwhile to note that this inconsistency exists despite Ma *et al.*'s 2005 work where he proved the optimal signal state MPN $\cong 0.5$ and the decoy state MPN $\cong 0.1$ [22]. Similarly, there is considerable disparity in the protocol occurrence percentages with signal states ranging from 50% to ~99%, decoy states ranging from <1% to 40%, and vacuum states ranging from 0% to 25%.

Despite the decoy state protocol's wide-spread employment, its effectiveness in detecting PNS attacks has not been thoroughly addressed in the literature. For example, in his defining work on the decoy state protocol, Lo states "Any attack by Eve that will change the value of any one of the $Y_n$'s and $e_n$'s substantially will, in principle, be caught with high probability by our decoy state method" [21]. Likewise, in the most detailed treatment available on the topic, the author merely states "significant deviation of the measured ratio from this expected value indicates a PNS by Eve" [43].

*A. Problem Formulation and Research Questions*

As the decoy state protocol is often employed in high performance QKD systems, and particularly the most impressive technology demonstrations to date (in terms of delivered key rate [26] and network size [27]), there is a need to understand its security effectiveness more fully. Moreover, it is important for system developers and users to understand how the protocol can be optimized to maximize both quantum throughput for secret key generation and detect PNS attacks (and variations thereof) with high confidence. Therefore, it is desirable to address the following research questions:

1) How do the signal and decoy state MPN values affect the system's ability to detect PNS attacks?

2) How does the difference between the signal and decoy state MPN values affect the system's ability to detect PNS attacks?

3) How do the signal, decoy, and vacuum state occurrence percentages affect the system's ability to detect PNS attacks?

4) How does variation in the generation and detection of signal and decoy states affect the system's ability to detect PNS attacks?

5) How does propagation distance (i.e., loss) affect the system's ability to differentiate between normal behavior and physical disturbances indicative of PNS attacks?

TABLE III
DECOY STATE ENABLED QKD SYSTEM CONFIGURATIONS

| Case | Signal MPN | Decoy MPN | Occurrence Percentage ($\mu / \nu / Y_0$) | Propagation Distance (km) | Key Rate (bps) |
|---|---|---|---|---|---|
| 1 [44] | 0.80 | 0.12 | 90 / 10 / 0 | 15 | 165 |
| 2 [45] | 0.55 | 0.152 | 63.5 / 20.3 / 16.2 | 60 | <428[*] |
| 3 [46] | 0.425 | 0.204 | 75 / 25 / 0[*] | 25 | 5.5k |
| 4 [47] | 0.6 | 0.2 | 50 / 40 / 10 | 75 | ~12 |
| 5 [47] | 0.6 | 0.2 | 50 / 40 / 10 | 102 | ~8 |
| 6 [48] | 0.487 | 0.064 | 83.1 / 12.3 / 4.6 | 85 | ~28 |
| 7 [48] | 0.297 | 0.099 | 83.1 / 12.3 / 4.6 | 100 | ~2 |
| 8 [49] | 0.27 | 0.39 | 87 / 9 / 4 | 144 | ~13 |
| 9 [50] | 0.55 | 0.098 | 93 / 6.2 / 1.6 | 20 | 10k |
| 10 [51] | 0.48 | 0.16 | 93 / 6.2 / 1.6 | 25 | 5.7k |
| 11 [52] | 0.55 | 0.10 | 80 / 16 / 4 | 20 | 1.02M |
| 12 [53] | 0.57 | 0.13 | 70 / 20 / 10 | 140 | ~2 |
| 13 [54] | 0.65 | 0.08 | 75 / 12.5 / 12.5 | 20 | 1.5k |
| 14 [54] | 0.60 | 0.20 | 75 / 12.5 / 12.5 | 20 | 1.6k |
| 15 [55] | 0.6 | 0.2 | 50 / 25 / 25 | 200 | 11.8 |
| 16 [56] | 0.6 | 0.2 | 50 / 25 / 25 | 200 | 15 |
| 17 [57] | 0.5 | 0.1 | 98.83 / 0.78 / 0.39 | 50 | 1.002M |
| 18 [58] | 0.6 | 0.2 | 75 / 12.5 / 12.5 | 8-60[**] | 1.2-4.5k[**] |
| 19 [27] | 0.65 | 0.1 | 87.5 / 6.25 / 6.25 | 30–80[**] | 0.8-16k[**] |
| 20 [26] | 0.4 | 0.04 | 98 / 1.5 / 0.5 | 45 | 300k |

[*] Value estimated or assumed from reference.
[**] Multiple systems employed.

*B. Experimental Design*

From the comprehensive listing of decoy state configurations captured in Table III, and detailed understanding of the decoy state protocol, five experimental factors are identified as shown in Table IV. First, operational distances of 20 and 50 km are selected to represent common metropolitan network lengths and long-haul backbone links. For those not familiar with quantum communication, losses of ~0.2 dB per km in single mode fiber significantly limit propagation distances where 20 km equates to 4 dB loss (or 40% efficiency) and 50 km equates to 10 dB loss (or 10% efficiency) [1]. Next, signal and decoy MPNs representative of normal and high configurations are chosen for examination. As the main focus of this study, five occurrence percentage configurations are selected for analysis. Lastly, each treatment is examined during normal conditions and when subject to PNS attacks. All other design and configuration settings are held constant (described in Section III-C).

TABLE IV
EXPERIMENTAL DESIGN

| Operational Distance | Signal MPN | Decoy MPN | Occurrence Percentage (Signal/Decoy/Vacuum) | PNS attack |
|---|---|---|---|---|
| 20 km | 0.5 | 0.1 | 60 / 30 / 10 | No |
| 50 km | 0.8 | 0.2 | 70 / 20 / 10 | Yes |
| | | | 80 / 10 / 10 | |
| | | | 90 / 5 / 5 | |
| | | | 99 / 0.5 / 0.5 | |

For this study a full factorial design was selected, as it is relatively easy to evaluate all 80 treatments in a simulation environment. In order to well characterize the system's behavior, and make statistically significant conclusions, 1,000 runs are executed for each treatment using the DoD's High Performance Computing Modernization Program at Wright-Patterson Air Force Base.

Regarding this experimental design, it is important to note that 20 km does not necessarily provide a sufficient loss budget for Eve to conduct PNS attacks without negatively impacting Bob's expected detection rate [59]. This is because Eve introduces loss on the quantum channel as she blocks all the single photon pulses sent by Alice. For example, Eve introduces ~7.4 dB loss against an MPN of 0.5, whereas the 20 km link only provides a ~4 dB loss budget for Eve to take advantage of with her lossless quantum teleportation channel. Despite this constraint, analyzing the decoy state protocol's ability to detect PNS attacks at this distance is desirable because many implementations have operational distances of 15-25 km as noted in Table III. Moreover, if Eve is able to insert herself on the quantum channel before protocol calibration, her presence would go unnoticed with respect to loss and key rate.

### C. Research Model

In this study, Alice is configured to generate signal, decoy, and vacuum pulses according to the decoy state protocol and BB84 polarization based prepare and measure protocol as described above. In particular, Alice is programmed to randomly prepare signal, decoy, and vacuum pulses according to the prescribed occurrence percentages at a 5 MHz pulse rate with commercially representative laser fluctuations. Alice then transmits the prepared pulses through the appropriate 20 or 50 km quantum channel, which has 4 or 10 dB loss respectively and induced physical disturbances. Bob's model includes 3.5 dB loss and representations of commercially available Avalanche Photo-Diode (APD) detectors each configured with 10% detector efficiency, a 5E-6 dark count rate (spontaneous detections when no photons are present), and a 0.01 after pulse rate (erroneous detections following a successful detection).

The research model was developed in a simulation framework specifically designed to capture and study the security and performance impact of implementation non-idealities in QKD systems, algorithms, and protocols [60]. For example, performance and security limitations with respect to speed, accuracy, and environmental disturbances are captured in the modeled laser source, decoy state generator, pulse modulator, quantum channel, and APD detectors. The decoy state enabled BB84 QKD model was developed in three increments each with increasing capability. The first increment provided a hardware-focused QKD notional architecture built in a modular fashion from a library of optical and electro-optical components [60]. The second increment added the processes and logic required to execute the decoy state protocol [61], [62]. In the third increment, modeled components were extended and the PNS attack was implemented [17]. Throughout model development, considerable effort was spent thoroughly defining, decomposing, modeling, verifying, and validating the decoy state enabled QKD model with each optical component verified against commercial specifications see Appendix of [60] and [63]. Additionally, the model was validated against eight fielded QKD systems [61] with additional modeling and simulation details presented in Section IV.

The research model allows analysts to uniquely study the security profile of the decoy state protocol (and other QKD protocols) in ways that are difficult or impossible with conventional means. For example, the model enables detailed analysis of the PNS attack which cannot yet be fully realized with current technologies [15], [16]. Moreover, the model uniquely enables detailed examination of each multiphoton pulse generated by Alice to determine if it is successfully split by Eve and detected by Bob. Thus, the researcher is able to explicitly know which pulses are compromised yet contribute to the QKD-generated secret key bits.

### IV. ANALYSIS OF RESULTS

In this section, the decoy state protocol's ability to detect PNS attacks is examined. First, the efficiency based method of detecting PNS attacks is explained, including expected operational variations from non-ideal optical components and processes. Next, simulation results for several common decoy state protocol configurations are described. Based on these results, an optimization of the decoy state protocol is presented and demonstrated. Lastly, implementation security guidance is offered for decoy state enabled QKD systems.

### A. Detecting PNS Attacks

Despite the creativeness of Eve's PNS attack, her detectability is based on the decoy state protocol's ability to differentiate between subtle changes in the signal and decoy states. In lieu of comparing photon number dependent yields $Y_n^{signal}$, $Y_n^{decoy}$ which are only measurable with expensive Photon Number Resolving (PNR) detectors [64], this study utilizes the efficiency based security condition [65]

$$\eta^{signal} = \eta^{decoy} \qquad (4)$$

where $\eta^{signal}$ is the signal state efficiency and $\eta^{decoy}$ is the decoy state efficiency. The efficiency based decoy state security method directly compares the signal and decoy state efficiencies from readily available measurements instead of requiring advanced technologies. The signal (and decoy) state efficiency is defined as

$$\eta^{signal} = \frac{-\ln|1 + Y_0 - Q_\mu|}{\mu} \quad (5)$$

where $Y_0$ is the system's measured dark count rate defined as

$$Y_0 = \frac{\text{Number of vacuum state detections}}{\text{Number of vacuum state pulses sent}} \quad (6)$$

$Q_\mu$ is the measured signal state gain defined as

$$Q_\mu = \frac{\text{Number of signal state detections}}{\text{Number of signal state pulses sent}} \quad (7)$$

and $\mu$ is the signal state's prescribed MPN (typically 0.5). This method also allows the QKD system to assure the quantum channel is free from unwanted attacks without *a priori* knowledge such as a well-characterized quantum channel as required in prior art.

*B. Expected Variation in the Decoy State Protocol*

Due to non-ideal devices, physical disturbances, and probabilistic single photon sources, variations are expected in the protocol's operation. These variations directly impact the system's ability to detect PNS attacks and must be accounted for, thus, the security condition becomes

$$\eta^{signal} = \eta^{decoy} \pm \Delta \quad (8)$$

where $\Delta$ represents the protocol's expected variation during quantum exchange. Variation in the decoy state efficiency is primarily considered because it exhibits significantly more variation than the signal state due to its reduced occurrence percentage and lower MPN.

While there are many potential sources of variation (e.g., fluctuations in laser sources, polarization dependent losses, variations in decoy state MPNs, temperature changes, physical disturbances, unstable detector efficiencies, etc.), many of them can be ignored due to the rapid propagation of photons through optical fiber (i.e., 2/3 the speed of light $\approx 2x10^8$ m/s). More explicitly, quantum exchange rounds (i.e., 100,000 signal state detections [66]) are typically very short (e.g., $< 20x10^{-3}$ s) and many of these effects are orders of magnitude slower (e.g., temperature change due to direct sunlight). Thus, Alice's pulse-to-pulse variation is of primary interest, and specifically, variation in her laser source (e.g., a commercially available id300 pulsed laser [67]) and decoy state generator (e.g., an electronically controlled Variable Optical Attenuator (VOA) used to control the MPN of each signal, decoy, and vacuum pulse [68]).

Fig. 3 illustrates Alice's modeled variation when calibrated to produce weak coherent optical pulses with an MPN of 0.55. Because of the large number of pulses, the 99.9% Prediction Interval (PI) characterizes her expected MPN variation well. This means, Alice will generate pulses with an MPN between 0.49 and 0.61 nearly 100% of the time. Thus, variations in generating signal, decoy, and vacuum pulses should be expected and addressed when considering the effectiveness of the decoy state protocol in detecting PNS attacks.

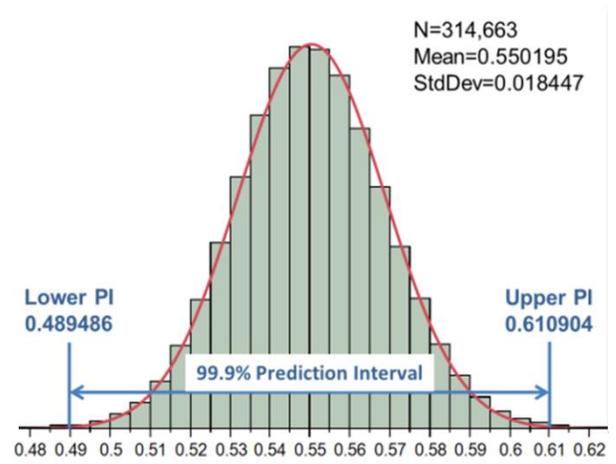

Fig. 3. Variation in weak coherent pulse MPN emitted from Alice due to laser fluctuations and decoy state performance limitations.

*C. Studying Detection Results*

Fig. 4 illustrates the normal operating conditions for 20 configurations over an operational distance of 20 km (the 50 km results are not shown because they are very similar). The results are grouped with respect to signal and decoy MPNs with each treatment labeled by signal-decoy-vacuum occurrence percentages. The overlapping box plots imply $\eta^{signal} = \eta^{decoy}$; thus, the system is operating in a secure state. Of note, variation in the signal state remains relatively fixed, while variation in the decoy state increases as the occurrence percentage lessens from 30% to 0.5%. Likewise, the lower MPN (i.e., 0.1 compared to 0.2) results in slightly more variation in each configuration. This occurs because less decoy states are sent by Alice, and therefore, detected by Bob, causing more variation. In all 40 configurations studied without PNS attacks at both 20 and 50 km, the signal and decoy state efficiencies are overlapping with no statistically significant differentiation.

Fig. 5 illustrates results over the 50 km operational distance from 20 configurations when subject to PNS attacks (the 20 km results are not shown because they are very similar). For each configuration studied, there is a clear separation between the decoy state efficiencies and the signal state efficiencies. This is because Eve inadvertently blocks most of the decoy state pulses since a majority of them contain only a single photon due to its lower MPN. Conversely, relatively few signal state pulses are blocked since the higher MPN generates more multi-photon pulses. Thus, Eve significantly reduces the decoy state efficiency and slightly elevates the signal state efficiency to compensate for Bob's expected detection rate. This behavior is precisely why the decoy state protocol requires two different MPNs in otherwise indistinguishable states (i.e., Eve is unaware of the pulse type, since any of the pulse states *could* consist of 0, 1, or ≥2 photons). Additionally, as can be seen in the downward trending efficiencies, these responses are tempered by the protocol's occurrence percentages and Eve's gain matching.

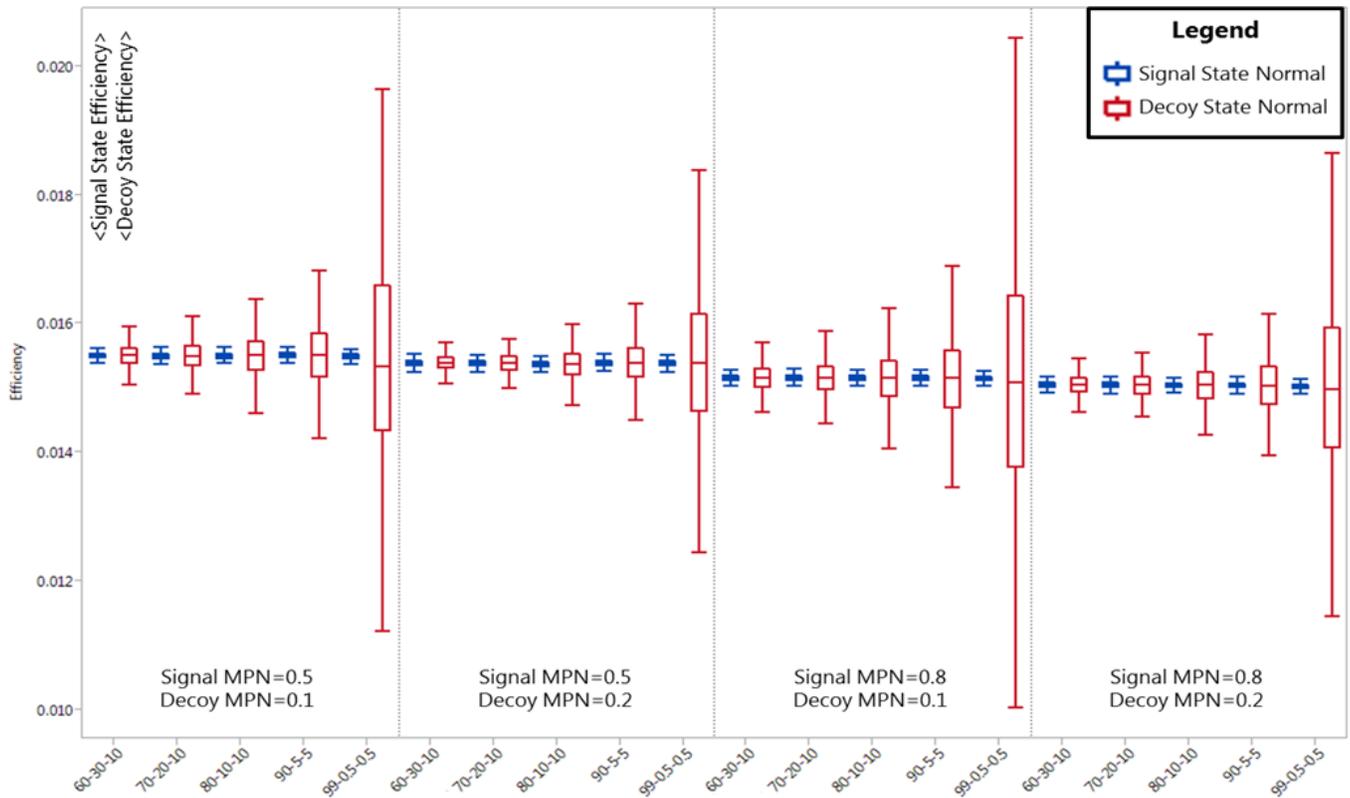

Fig. 4. Simulation results are shown for the for the 20 km decoy state protocol configurations examined when operating under normal conditions. In each configuration studied, the signal and decoy state efficiencies are the same $\eta^{signal} = \eta^{decoy} \pm \Delta$ (within expected variation tolerances).

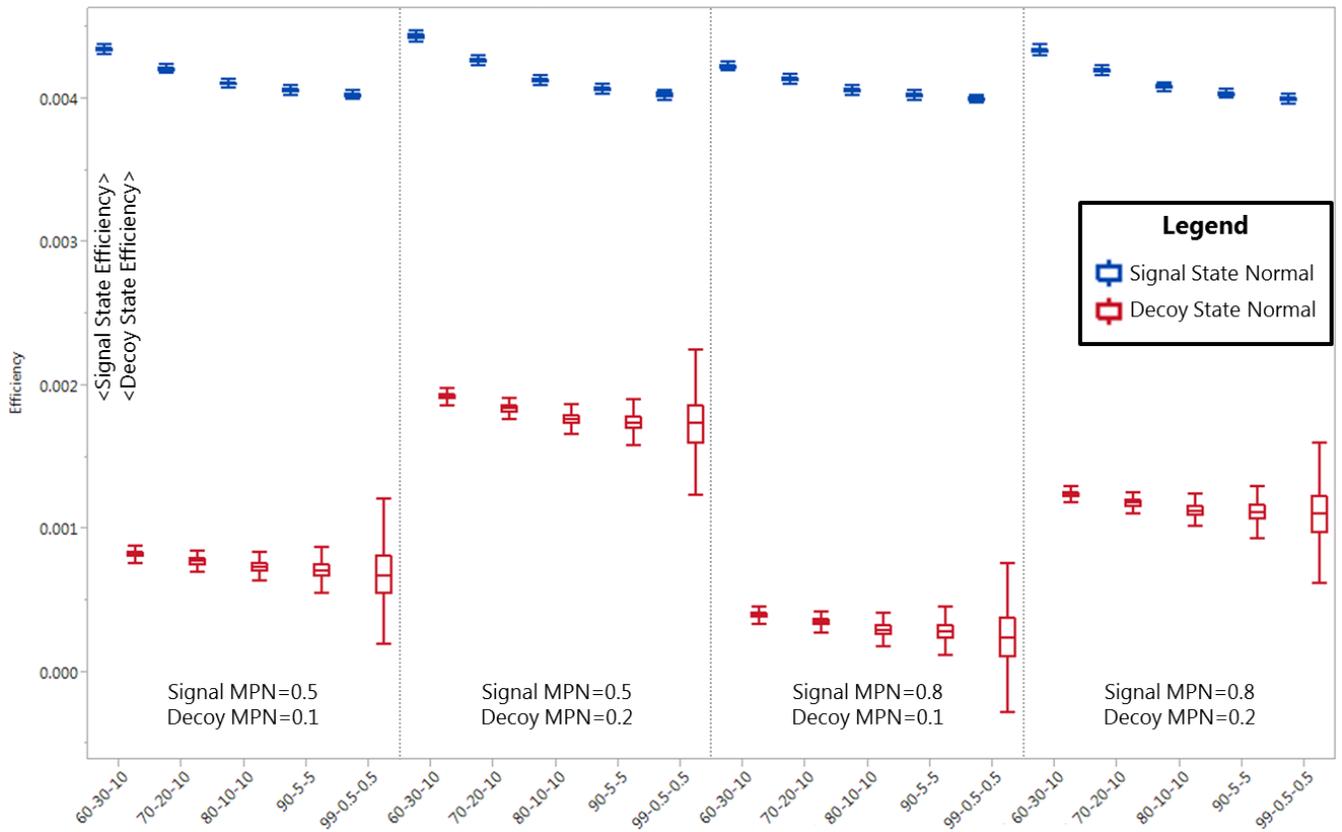

Fig. 5. Simulation results are shown for the 50 km decoy state protocol configurations examined when subject to PNS attacks. In each configuration studied, the signal and decoy state efficiencies are statistically different $\eta^{signal} \neq \eta^{decoy} \pm \Delta$ (outside expected variation tolerances).

In the 40 configurations considered at both 20 and 50 km distances, the PNS attack was successfully detected in all 40,000 trials (i.e., 1,000 trials in each of the 40 configurations studied). For example, in the worst case scenario, when the signal and decoy state MPNs are closest (0.5 and 0.2) with the least amount of decoy states (99% signal, 0.5% decoy, and 0.5% vacuum) and the most loss (10 dB loss over the 50 km channel), there is very strong statistical evidence that $\eta^{signal} \neq \eta^{decoy} \pm \Delta$ with $P < 0.001$. These results demonstrate the decoy state protocol's ability to detect PNS attacks across a wide set of commonly implemented configurations, and perhaps more importantly, they demonstrate that the protocol can be further optimized as identified by the large "white space" between the signal and decoy states efficiencies in Fig. 5.

### D. Optimization for Performance and Security

While the decoy state protocol has been optimized with respect to MPNs contributing to secret key distribution [22], the protocol has not been optimized for detecting PNS attacks. Hence, we provide an optimization which assures high security confidence and allows the protocol's performance to be maximized based on a detailed study of signal and decoy state MPNs and occurrence percentages, as well as, design decisions and architectural considerations.

From this study we learn that the protocol's ability to detect PNS attacks is primarily controlled by losses due to each state's occurrence percentage, MPN, and the end-to-end quantum communication path. More specifically, to detect PNS attacks in real-time with high confidence only a few decoy state detections are necessary during each round of quantum exchange (i.e., a predetermined number of detections). For example, the decoy state protocol can be configured to perform the PNS attack check after each round of 100,000 detections. Furthermore, we learn that an arbitrarily high level of confidence (e.g., >99.9%) is possible because statistical confidence is increased through multiple rounds of quantum exchange and not the number of decoy state detections per round.

In order to optimize the decoy state protocol, the developer should choose the highest signal state occurrence percentage possible, while meeting the minimum number of decoy state detections to reliably detect PNS attacks (i.e., choose the minimal decoy occurrence percentage possible). Assuming the suggested MPNs of Ma *et al.* are used ($\mu = 0.5, \nu = 0.1$) [22], the optimized decoy state protocol configuration can be described in a system of equations. First, the signal state occurrence percentage $S_\mu$ should be as a close to unity as possible

$$S_\mu \to 1 \tag{9}$$

where $S_\mu$ is limited by the decoy and vacuum state occurrence percentages $S_\nu$, $S_{Y_0}$, respectively

$$S_\mu = 1 - S_\nu - S_{Y_0}. \tag{10}$$

Accordingly, it is advantageous to minimize both $S_\nu$ and $S_{Y_0}$; however, the decoy state occurrence percentage $S_\nu$ must be high enough to effectively differentiate between noise on the quantum channel and a PNS attack where the decoy state gain $Q_\nu$ must exceed the system's measured dark count rate $Y_0$

$$Q_\nu > Y_0. \tag{11}$$

This condition implies at least one decoy state detection $N_\nu$ per round of quantum exchange which is not due to a dark count (i.e., a signal to noise ratio >1).

Thus, the optimized decoy state configuration can be further clarified

$$S_\nu \ll 1 \tag{12}$$

$$N_\nu \geq 1. \tag{13}$$

For a given architecture, the optimized decoy state protocol can be determined from the minimum number of decoy state detections $N_\nu$, the desired number of signal state detections $N_\mu$, the signal and decoy state gains $Q_\mu$, $Q_\nu$, and their occurrence percentages $S_\mu$, $S_\nu$ where

$$N_\nu = S_\nu Q_\nu N_{total\ pulses\ sent} \tag{14}$$

$$N_{total\ pulses\ sent} = \frac{N_\mu}{S_\mu Q_\mu} \tag{15}$$

$$S_\nu = \frac{N_\nu S_\mu Q_\mu}{Q_\nu N_\mu}. \tag{16}$$

While the necessary parameters for optimization are readily available, in order maximize performance the system's architecture must be well-characterized in the desired operational environment. This is because the decoy state protocol is being configured to operate at its minimum threshold and is extremely sensitive to implementation non-idealities and performance variations to include Alice's ability to generate weak coherent pulses, losses in the quantum channel, physical disturbances, detector efficiency, and particularly the system's operational dark count rate.

### E. Example Optimization

In this section, an optimization of a fielded decoy state enabled QKD system is demonstrated. As one of the most well documented decoy state protocol implementations and a major milestone in the world's largest QKD network, Chen *et al.*'s work lends itself well to detailed analysis [54]. The protocol's configuration is provided in Table V.

TABLE V
DECOY STATE PROTOCOL IMPLEMENTATION [58]

| Protocol Configuration | Operational Results |
|---|---|
| $S_\mu = 0.75$ | $\eta = 0.00985$ |
| $S_\nu = 0.125$ | $Q_\mu = 6.36\text{E-}3$ |
| $S_{Y_0} = 0.125$ | $Q_\nu = 8.61\text{E-}4$ |
| $\mu = 0.65$ | $Y_0 = 1.0\text{E-}4$ |
| $\nu = 0.08$ | |

Assuming $N_\mu = 100{,}000$ detections per quantum exchange and an arbitrarily small vacuum state occurrence percentage $S_{Y_0} = 0.005$, the decoy state protocol occurrence percentages can be optimized to $S_\mu = 0.99435$, $S_v = 0.00065$ using the approach described in Eqs. (9-16). This optimized configuration is particularly advantageous as it results in a >30% increase in key distribution (i.e., a signal state occurrence percentage 99.435% instead of 75%) and the ability to detect PNS attack with 99.9% confidence at no additional cost.

Fig. 6 presents detailed results of the optimized protocol while operating under normal conditions and when subject to PNS attacks. Shown on the left, during normal operations the signal and decoy state efficiencies (blue and red) overlap as expected. Shown in the middle, PNS attacks cause the signal and decoy state efficiencies (green and purple) to become non-overlapping. In particular, since the protocol is configured to operate with a minimum number of decoy state detections, the PNS attack reduces the decoy state from a small number of detections to zero during nearly every round of quantum exchange. This results in a reported decoy state mean efficiency of 0.000 with relatively little variation (see Fig. 7 for further details). Consequently, the optimized decoy state protocol configuration serves to emphasize the negative impact of the PNS attack by forcing the decoy state's efficiency below the measured dark count rate (shown in brown with a detailed inlay) because so few decoy state detections are expected per round of quantum exchange.

During normal operations, the optimized configuration results in at least one decoy state detection per 100,000 detections and a mean of 9 detections. Conversely, only a few decoy state detections are expected during PNS attacks; however, they are statistical outliers occurring in only 134 out of the 1,000 rounds of quantum exchange. In terms of efficiency, the mean decoy state efficiency is 0.0096 during normal operations and drops to 0.0013 during PNS attacks. As a result, the PNS attack is readily detectable with a high statistical confidence of $P < 0.001$ when considering 1,000 rounds of quantum exchange with 100,000 detections per round.

While the decoy state occurrence percentage $S_v$ can be further reduced, statistical significance begins to diminish because the number of decoy state detections per round of quantum exchange approaches zero during normal operations. Moreover, as the occurrence percentage is further reduced the protocol's integrity is jeopardized as the decoy state gain must be larger than the system's dark count rate (i.e., $Q_v > Y_0$).

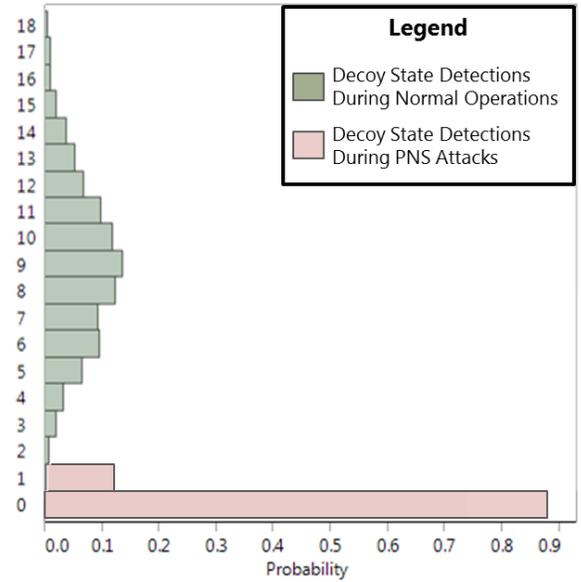

Fig. 7. Simulation results detailing the number of decoy state detections per round of quantum exchange for the optimized decoy state protocol based on the fielded QKD system in [54].

### F. Implementation Recommendations

In addition to the protocol optimization described above and the findings therein, this research effort led to several design and implementation recommendations for commercially viable QKD systems:

1) Upon system startup, the decoy state protocol should be configured to quickly perform initial security checks to ensure the quantum channel is free from PNS attacks. For example, 1,000 rounds of quantum exchange can be executed in a relatively short amount of time during initial calibration activities.

2) Configure the decoy state protocol to continuously monitor for PNS attacks in real-time and over several rounds of quantum exchange to increase confidence in the system's security.

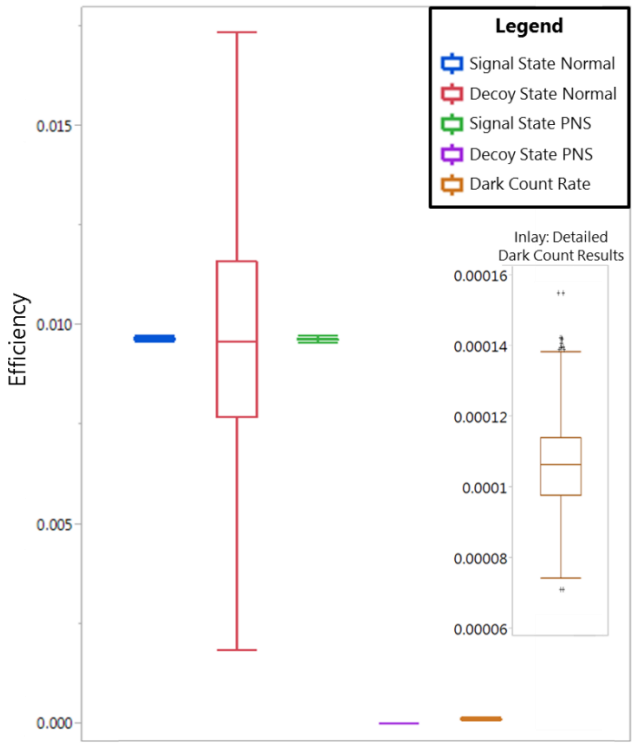

Fig. 6. Simulation results show the optimized decoy state protocol for detecting PNS attacks based on the fielded QKD system [54].

Fig. 7 displays the number of decoy state detections per round of quantum exchange during normal operations (shown in green) and when subject to PNS attacks (shown in red).

3) The noise level (i.e., the dark count rate) should be measured during dedicated calibration activities with very large numbers of vacuum signals (e.g., $\geq 10^9$) intermixed with signal and decoy states to well-characterize the operational environment and system architecture.

4) During operation, the dark count rate should be compared to the calibration results in order to detect changes in the operational environment such as temperature changes or additional physical disturbances.

5) Minimize the vacuum state occurrence percentage but do not eliminate it. The state it can be used as an indicator to monitor for attacks such as the blinding attack [69].

Additionally, while Ma *et al.*'s work optimized the signal state MPN at ~0.5, users may want to consider higher signal state MPNs such as those successfully demonstrated in the world's largest QKD network (i.e., $\mu = 0.65$) [27]. Moreover, past work on the subject recommends MPNs on the order of 1.0 - 1.2 based on pragmatic technical assumptions [70].

V. CONCLUSIONS

In this study, the ability of the decoy state enabled QKD systems to detect PNS attacks is analyzed and demonstrated. In contrast to most decoy state protocol research, this work focuses on the optimization of the protocol's occurrence percentages to both maximize secret key rates and assure PNS attacks are detectable with high confidence. Additionally, practical implementation performance and security guidance is provided for system developers and users. Lastly, this work demonstrates a repeatable methodology for studying QKD systems to support formal certification efforts [71].

Future suggested work includes optimization of the decoy state protocol in a QKD system for validation, and perhaps, tested against a PNS or PNS-like attack as suggested in [16].

DISCLAIMER

References

[1] N. Gisin, G. Ribordy, W. Tittel and H. Zbinden, "Quantum cryptography," *Reviews of Modern Physics,* vol. 74, no. 1, pp. 145-195, 2002.

[2] V. Scarani and C. Kurtsiefer, "The black paper of quantum cryptography: real implementation problems," *arXiv:0906.4547v2,* 2009.

[3] L. Oesterling, D. Hayford and G. Friend, "Comparison of commercial and next generation quantum key distribution: Technologies for secure communication of information," in *Homeland Security (HST), 2012 IEEE Conference on Technologies for,* 2012.

[4] V. Scarani, H. Bechmann-Pasquinucci, N. J. Cerf, M. Dušek, N. Lütkenhaus and M. Peev, "The security of practical quantum key distribution," *Reviews of Modern Physics,* vol. 81, no. 3, pp. 1301-1350, 2009.

[5] L. O. Mailloux, M. R. Grimaila, D. D. Hodson, G. Baumgartner and C. McLaughlin, "Performance evaluations of quantum key distribution system architectures," *IEEE Security and Privacy,* vol. 13, no. 1, pp. 30-40, 2015.

[6] C. Elliott, "Quantum cryptography," *IEEE Security & Privacy,* vol. 2, no. 4, pp. 57-61, 2004.

[7] B. Qi, L. Qian and H.-K. Lo, "A brief introduction of quantum cryptography for engineers," *arXiv: 1002.1237,* 2010.

[8] S. Wiesner, "Conjugate coding," *ACM Sigact News,* vol. 15, no. 1, pp. 78-88, 1983.

[9] C. H. Bennett and G. Brassard, "Quantum cryptography: public key distribution and coin tossing," in *Proceedings of IEEE International Conference on Computers, Systems and Signal Processing,* 1984.

[10] G. S. Vernam, "Cipher printing telegraph systems for secret wire and radio telegraphic communications," *American Institute of Electrical Engineers, Transactions of the,* vol. 45, pp. 295-301, 1926.

[11] C. E. Shannon, "Communication theory of secrecy systems," *Bell System Technical Journal,* vol. 28, pp. 656-715, 1949.

[12] Quantum Cryptography Conference, "QCrypt 2015," 2016. [Online]. Available: 2015.qcrypt.net.

[13] D. Gottesman, H.-K. Lo, N. Lutkenhaus and J. Preskill, "Security of quantum key distribution with imperfect devices," in *In Information Theory, 2004. ISIT 2004. Proceedings. International Symposium on*, 2004.

[14] R. Renner, N. Gisin and B. Kraus, "An information-theoretic security proof for QKD protocols," *Physical Review A,* vol. 72, no. 1, p. 012332, 2005.

[15] G. Brassard, N. Lutkenhaus, T. Mor and B. C. Sanders, "Limitations on practical quantum cryptography," *Physical Review Letters,* vol. 85, no. 6, p. 1330, 2000.

[16] N. Lütkenhaus, "Security against individual attacks for realistic quantum key distribution," *Physical Review A,* vol. 61, no. 5, p. 052304, 2000.

[17] L. Mailloux, D. Hodson, M. Grimaila, R. Engle, C. McLaughlin and G. Baumgartner, "Using modeling and simulation to study photon number splitting attacks," *IEEE Access,* vol. 4, pp. 2188-2197, 2016.

[18] S. Loepp and W. K. Wootters, Protecting Information, New York: Cambridge University Press, 2006.

[19] G. Nogues, A. Rauschenbeutel, S. Osnaghi, M. Brune, J. M. Raimond and S. Haroche, "Seeing a single photon without destroying it," *Nature,* vol. 400, no. 6741, pp. 239-242, 1999.

[20] W.-Y. Hwang, "Quantum key distribution with high loss: toward global secure communication," *Physical Review Letters,* vol. 91, no. 5, p. 057901, 2003.

[21] H.-K. Lo, X. Ma and K. Chen, "Decoy state quantum key distribution," *Physical Review Letters,* vol. 94, no. 3, p. 230504, 2005.

[22] X. Ma, B. Qi, Y. Zhao and H.-K. Lo, "Practical decoy state for quantum key distribution," *Physical Review,* vol. 72, no. 1, p. 012326, 2005.

[23] X.-B. Wang, "Beating the photon-number-splitting attack in practical quantum cryptography," *Physical Review Letters,* vol. 94, no. 23, p. 230503, 2005.

[24] X.-B. Wang, "Decoy-state protocol for quantum cryptography with four different intensities of coherent light," *Physical Review A,* vol. 72, no. 1, p. 012322, 2005.

[25] J. W. Harrington, J. M. Ettinger, R. J. Hughes and J. E. Nordholt, "Enhancing practical security of quantum key distribution with a few decoy states," *arXiv,* pp. quant-ph/0503002, 2005.

[26] A. R. Dixon, J. F. Dynes, M. Lucamarini, B. Fröhlich, A. W. Sharpe, A. Plews, S. Tam and e. al., "High speed prototype quantum key distribution system and long term field trial," *Optics Express,* vol. 23, no. 6, pp. 7583-7592, 2015.

[27] S. Wang, W. Chen, Z.-Q. Yin, H.-W. Li, D.-Y. He, Y.-H. Li, Z. Zhou and e. al., "Field and long-term demonstration of a wide area quantum key distribution network," *Optics Express,* vol. 22, no. 18, pp. 21739-21756, 2014.

[28] X.-B. Wang, C.-Z. Peng and J.-W. Pan, "Simple protocol for secure decoy-state quantum key distribution with a loosely controlled source," *arXiv:quant-ph/0609137,* 2006.

[29] X.-B. Wang, "Secure and efficient decoy-state quantum key distribution with inexact pulse intensities," *arXiv:quant-ph/0609081,* 2006.

[30] W. Mauerer and C. Silberhorn, "Quantum key distribution with passive decoy state selection," *Physical Review A,* vol. 75, no. 5, p. 050305, 2007.

[31] M. Hayashi, "General theory for decoy-state quantum key distribution with an arbitrary number of intensities," *New Journal of Physics,* vol. 9, no. 8, p. 284, 2007.

[32] X.-B. Wang, C.-Z. Peng and J.-W. Pan, "Simple protocol for secure decoy-state quantum key distribution with a loosely controlled source," *Applied Physics Letters,* vol. 90, no. 3, p. 031110, 2007.

[33] T. Tsurumaru, A. Soujaeff and S. Takeuchi, "Exact minimum and maximum of yield with a finite number of decoy light intensities," *Physical Review A,* vol. 77, no. 2, p. 022319, 2008.

[34] X.-B. Wang, L. Yang, C.-Z. Peng and J.-W. Pan, "Decoy-state quantum key distribution with both source errors and statistical fluctuations," *New Journal of Physics,* vol. 11, no. 7, p. 075006, 2009.

[35] J.-Z. Hu and X.-B. Wang, "Reexamination of the decoy-state quantum key distribution with an unstable source," *Physical Review A,* vol. 82, no. 1, p. 012331, 2010.

[36] J.-Z. Hu and X.-B. Wang, "Secure quantum key distribution in an easy way," *arXiv:1004.3730,* 2010.

[37] B. W.-s. Yuan Li, H.-w. Li, C. Zhou and Y. Wang, "Passive decoy-state quantum key distribution for the weak coherent photon source with intensity fluctuations," *arXiv:1312.7383 [quant-ph],* 2013.

[38] B. W.-s. Yuan Li, H.-w. Li, C. Zhou and Y. Wang, "Passive decoy-state quantum key distribution for the weak coherent photon source with intensity fluctuations," *arXiv:1312.7383 [quant-ph],* 2013.

[39] Q.-C. Sun, W.-L. Wang, Y. Liu, F. Zhou, J. Pelc, M. M. Fejer, C.-Z. Peng and e. al., "Experimental passive decoy-state quantum key distribution," *Laser Physics Letters,* vol. 11, no. 8, p. 085202, 2014.

[40] J. Hasegawa, M. Hayashi, T. Hiroshima, A. Tanaka and A. Tomita, "Experimental decoy state quantum key distribution with unconditional security incorporating finite statistics," *arXiv:0705.3081,* 2007.

[41] M. Lucamarini, K. A. Patel, J. F. Dynes, B. Fröhlich, A. W. Sharpe, A. R. Dixon, Z. L. Yuan, R. V. Penty and A. J. Shields, "Efficient decoy-state quantum key distribution with quantified security," *Optics express,* vol. 21, no. 21, pp. 24550-24565, 2013.

[42] C. C. W. Lim, M. Curty, N. Walenta, F. Xu and H. Zbinden, "Concise security bounds for practical decoy-state quantum key distribution," *Physical Review A,* vol. 89, no. 2, p. 022307, 2014.

[43] Z. L. Yuan, A. W. Sharpe and A. J. Shields, "Unconditionally secure one-way quantum key distribution using decoy pulses," *Applied physics letters,* vol. 90, no. 1, p. 011118, 2007.

[44] Y. Zhao, B. Qi, X. Ma, H.-K. Lo and L. Qian, "Experimental quantum key distribution with decoy states," *Physical Review Letters,* vol. 96, no. 7, p. 070502, 2006.

[45] Y. Zhao, B. Qi, X. Ma, H.-K. Lo and L. Qian, "Simulation and implementation of decoy state quantum key distribution over 60km telecom fiber," in *Information Theory, 2006 IEEE International Symposium on,* 2006.

[46] Z. L. Yuan, A. W. Sharpe and A. J. Shields, "Unconditionally secure one-way quantum key distribution using decoy pulses," *arXiv:quant-ph/0610015,* 2006.

[47] C.-Z. Peng, J. Zhang, D. Yang, W.-B. Gao, H.-X. Ma, H. Yin, H.-P. Zeng, T. Yang, X.-B. Wang and J.-W. Pan, "Experimental long-distance decoy-state quantum key distribution based on polarization encoding," *Physical Review Letters,* vol. 98, no. 1, p. 010505, 2007.

[48] D. Rosenberg, J. W. Harrington, P. R. Rice, P. A. Hiskett, C. G. Peterson, R. J. Hughes, A. E. Lita, S. W. Nam and J. E. Nordholt, "Long-distance decoy-state quantum key distribution in optical fiber," *Physical Review Letters,* vol. 98, no. 1, p. 010503, 2007.

[49] T. Schmitt-Manderbach, H. Weier, M. Fürst, R. Ursin, F. Tiefenbacher, T. Scheidl, J. Perdigues and e. al., "Experimental demonstration of free-space decoy-state quantum key distribution over 144 km," *Physical Review Letters,* vol. 98, no. 1, p. 010504, 2007.

[50] J. F. Dynes, Z. L. Yuan, A. W. Sharpe and A. J. Shields, "Practical quantum key distribution over 60 hours at an optical fiber distance of 20km using weak and vacuum decoy pulses for enhanced security," *Optics Express,* vol. 15, no. 13, pp. 8465-8471, 2007.

[51] J. F. Dynes, Z. L. Yuan, A. W. Sharpe and A. J. Shields, "Decoy pulse quantum key distribution for practical purposes," *Optoelectronics, IET,* vol. 2, no. 5, pp. 195-200, 2008.

[52] A. R. Dixon, Z. L. Yuan, J. F. Dynes, A. W. Sharpe and A. J. Shield, "Gigahertz decoy quantum key distribution with 1 Mbit/s secure key rate," *Optics Express,* vol. 16, no. 23, pp. 18790-18979, 2008.

[53] D. Rosenberg, C. G. Peterson, J. W. Harrington, P. R. Rice, N. Dallmann, K. T. Tyagi, K. P. McCabe and e. al., "Practical long-distance quantum key distribution system using decoy levels," *New Journal of Physics,* vol. 11, no. 4, p. 045009, 2009.

[54] T.-Y. Chen, H. Liang, Y. Liu, W.-Q. Cai, L. Ju, W.-Y. Liu, J. Wang, H. Yin, K. Chen, Z.-B. Chen, C.-Z. Peng and J.-W. Pan, "Field test of a practical secure communication network with decoy-state quantum cryptography," *Optics Express,* vol. 17, no. 8, pp. 6540-6549, 2009.

[55] T.-Y. Chen, J. Wang, Y. Liu, W.-Q. Cai, X. Wan, L.-K. Chen, J.-H. Wang and e. al., "200km Decoy-state quantum key distribution with photon polarization," *arXiv,* p. arXiv:0908.4063, 2009.

[56] Y. Liu, T.-Y. Chen, J. Wang, W.-Q. Cai, X. Wan, L.-K. Chen, J.-H. Wang and e. al., "Decoy-state quantum key distribution with polarized photons over 200 km," *Optics Express,* vol. 18, no. 8, pp. 8587-8594, 2010.

[57] A. R. Dixon, Z. L. Yuan, J. F. Dynes, A. W. Sharpe and A. J. Shields, "Continuous operation of high bit rate quantum key distribution," *Applied Physics Letters,* vol. 96, no. 16, p. 161102, 2010.

[58] T. Chen, J. Wang, H. Liang, W. Liu, Y. Liu, X. Jiang, Y. Wang, X. Wan, W. Cai, L. Ju and L. Chen, "Metropolitan all-pass and inter-city quantum communication network," *Optics Express,* vol. 18, no. 26, pp. 27217-27225, 2010.

[59] N. Lütkenhaus and M. Jahma, "Quantum key distribution with realistic states: photon-number statistics in the photon-number splitting attack," *New Journal of Physics,* vol. 4, no. 1, p. 44.1–44.9, 2002.

[60] L. O. Mailloux, J. D. Morris, M. R. Grimaila, D. D. Hodson, D. R. Jacques, J. M. Colombi, C. McLaughlin, R. Engle and J. Holes, "A modeling framework for studying quantum key distribution system implementation non-idealities," *IEEE Access,* vol. 3, pp. 110-130, 2015.

[61] L. O. Mailloux, R. D. Engle, M. R. Grimaila, D. D. Hodson and C. McLaughlin, "Modeling decoy state quantum key distribution systems," *The Journal of Defense Modeling and Simulation: Applications, Methodology, Technology,* vol. 12, no. 4, pp. 489-506, 2015.

[62] R. Engle, M. Grimaila, L. Mailloux, D. Hodson, C. McLaughlin and G. Baumgartner, "Implementing the decoy state protocol in a practically-oriented quantum key distribution system-level model," *The Journal of Defense Modeling and Simulation: Applications, Methodology, Technology,* Submitted 2016.

[63] J. Holes, L. Mailloux, M. Grimaila and D. Hodson, "An Efficient Testing Process for a Quantum Key Distribution System Modeling Framework," in *International Conference on Scientific Computing (CSC15),* Las Vegas, NV, July 27-30, 2015, 2014.

[64] R. H. Hadfield, "Single-photon detectors for optical quantum information applications," *Nature photonics,* vol. 3, no. 12, pp. 696-705, 2009.

[65] L. O. Mailloux, M. R. Grimaila, J. M. Colombi, D. D. Hodson, R. D. Engle, C. V. McLaughlin and G. Baumgartner, "Quantum key distribution: examination of the decoy state protocol," *IEEE Communications Magazine,* vol. 53, no. 10, pp. 24-31, 2015.

[66] A. Mink and A. Nakassis, "LDPC for qkd reconciliation," *The Computing Science and Technology International Journal,* vol. 2, no. 2, pp. 6-14, 2012.

[67] ID Quantique, "id300 Series Sub-Nanosecond Pulsed Laser Source Datasheet," 2012. [Online]. Available: http://www.idquantique.com/images/stories/PDF/id300-laser-source/id300-specs.pdf. [Accessed 05 Mar 2014].

[68] OPLINK, "Electronically Variable Optical Attenuators," 2014. [Online]. Available: http://www.oplink.com/pdf/EVOA-S0012.pdf .

[69] L. Lydersen, C. Wiechers, C. Wittmann, D. Elser, J. Skaar and V. Makarov, "Hacking commercial quantum cryptography systems by tailored bright illumination," *Nature Photonics,* vol. 4, no. 10, pp. 686-689, 2010.

[70] D. Pearson and C. Elliott, "On the optimal mean photon number for quantum cryptography," *quant-ph/0403065,* 2004.

[71] ETSI, "Quantum key distribution," 08 June 2015. [Online]. Available: www.etsi.org/technologies-clusters/technologies/quantum-key-distribution.